\documentclass[letterpaper, 10 pt, conference]{ieeeconf} 
\IEEEoverridecommandlockouts                              
\usepackage{xcolor}
\usepackage{tikz}
\usetikzlibrary{positioning}
\usetikzlibrary{arrows}
\usetikzlibrary{circuits.ee.IEC}
\usepackage{dblfloatfix}
\usepackage{cite}
\usepackage{amssymb,amsfonts}
\usepackage{amsmath}
\usepackage{booktabs}
\usepackage[font=footnotesize]{caption}
\usepackage[font=footnotesize]{subcaption}
\usepackage[bottom]{footmisc}
\usepackage[english]{babel}
\usepackage{makecell}

\usepackage[capitalize,nameinlink,compress]{cleveref}
\crefname{figure}{Fig.}{Figures} 
\crefname{line}{line}{lines} 
\crefname{claim}{Claim}{Claims} 
\crefname{equation}{}{} 
\crefname{problem}{Problem}{Problems}
\crefname{assumption}{Assumption}{Assumptions}

\usepackage{amsthm}

\newtheorem{remark}{Remark}

\definecolor{backgroundcolor}{rgb}{0.4660, 0.6740, 0.1880}
\definecolor{backgroundcolor2}{rgb}{0, 0.4470, 0.7410}

\title{\LARGE \bf
MIMO Grid Impedance Identification of Three-Phase Power Systems: Parametric vs. Nonparametric Approaches
}

\author{Verena Häberle, Linbin Huang, Xiuqiang He, Eduardo Prieto-Araujo, Roy S. Smith, Florian Dörfler
\thanks{This work was supported by the European Union's Horizon 2020 and 2023 research and innovation programs (Grant Agreement Numbers 883985 and 101096197). Verena Häberle, Linbin Huang, Xiuqiang He, Roy Smith and Florian Dörfler are with the Automatic Control Laboratory, ETH Zurich, 8092 Zurich, Switzerland. Email:\{verenhae,linhuang,xiuqhe,rsmith,dorfler\}@ethz.ch. Eduardo Prieto-Araujo is a Serra Húnter Lecturer with CITCEA, Universitat Politècnica de Catalunya, 08028 Barcelona, Spain. Email: eduardo.prieto-araujo@upc.edu}%
}

\begin{document}

\maketitle
\thispagestyle{empty}
\pagestyle{empty}

\begin{abstract}
A fast and accurate grid impedance measurement of three-phase power systems is crucial for online assessment of power system stability and adaptive control of grid-connected converters. Existing grid impedance measurement approaches typically rely on pointwise sinusoidal injections or sequential wideband perturbations to identify a \textit{nonparametric} grid impedance curve via fast Fourier computations in the frequency domain. This is not only time-consuming, but also inaccurate during time-varying grid conditions, while on top of that, the identified nonparametric model cannot be immediately used for stability analysis or control design. To tackle these problems, we propose to use \textit{parametric} system identification techniques (e.g., prediction error or subspace methods) to obtain a parametric impedance model directly from time-domain current and voltage data. Our approach relies on injecting wideband excitation signals in the converter's controller and allows to accurately identify the grid impedance in closed loop within one injection and measurement cycle. Even though the underlying parametric system identification techniques are well-studied in general, their utilization in a grid impedance identification setup poses specific challenges, is vastly underexplored, and has not gained adequate attention in urgent and timely power systems applications. To this end, we demonstrate in numerical experiments how the proposed parametric approach can accomplish a significant improvement compared to prevalent nonparametric methods.
\end{abstract}

\section{Introduction}
In an effort to make the future electric power system more sustainable, a large-scale and increased utilization of power-electronics-based technologies in the grid is happening all around the world. Consequently, the power system dynamics have started to change and new stability problems are emerging. For instance, the interaction between power converters and the grid can lead to degradation of the power quality or even cause instabilities \cite{cespedes2009renewable,li2017unstable,wang2018harmonic}. 

Impedance-based analysis is considered as a powerful tool to assess the small-signal stability of grid-connected converters (e.g., by using the impedance-based stability criterion in~\cite{sun2011impedance}) and to adaptively change the control strategy for stability improvement\cite{sun2009small,sun2011impedance,cespedes2012online}. To do so, the small-signal dynamics of the power grid are modelled as a dynamic impedance equivalent in the form of a multi-input multi-output (MIMO) transfer function $Z_\mathrm{g}(s)$, which relates the terminal voltage $v$ and current $i$ at the point of common coupling (PCC) of a grid-connected converter system (\cref{fig:scheme}). Since in practice, however, the power grid is ever-changing and usually unknown from the perspective of converters, it is favorable to perform online grid impedance measurements to enable real-time impedance-based analysis and control.

Several methods for grid impedance identification/mea- surement have been proposed in literature so far; see~\cref{fig:mindmap} for a bird's eye overview. The most prevalent method for real-world applications is the frequency sweep method \cite{francis2011algorithm,huang2009small}, where the grid impedance is measured “point-by-point”, by injecting sinusoidal signals with different frequencies to obtain the whole impedance curve in the frequency domain. This, however, requires an extra device for the sinusoidal injection and is additionally highly time-consuming since the excitation injection process needs to be repeated many times. Furthermore, versatile renewable energy sources may change the grid operating point during the long-lasting impedance measurement, which can lead to inaccurate results.
\begin{figure}[t]
    \centering
    \usetikzlibrary{circuits.ee.IEC}
\usetikzlibrary{arrows}

\resizebox {0.4\textwidth} {!} {
\tikzstyle{roundnode} =[circle, draw=blue!60, fill=blue!5, scale = 0.5]
\begin{tikzpicture}[circuit ee IEC,scale=0.47, every node/.style={scale=0.65}]

\node [color=backgroundcolor] at (12.1,4) {$Z_\mathrm{g}(s)$};

\draw  plot[smooth, tension=.7] coordinates {(14.8,3.5) (13.7,3.2)(13.2,2.5) (13.7,1.3) (14.7,0.3)};
\node at (14.2,2.5) {utility};
\node at (14.2,2) {grid};

\draw (11.3,2.6)  -- (7.3,2.6) -- (13.2,2.6) node (v5) {};
\draw  (7.3,3.2) rectangle (3.4,1.9);
\node at (5.35,2.8) {grid-connected};
\node at (5.35,2.3) {converter system};

\draw[color=backgroundcolor]  (8.6,2.6) ellipse (0.05 and 0.15);
\draw [-latex,color =backgroundcolor](8.6,2.45) -- (8.6,1.6);
\draw[-latex,color=backgroundcolor] (9.7,2.6) -- (9.7,1.6);
\fill[color=backgroundcolor] (9.7,2.6) circle(0.6mm);
\draw [rounded corners = 4,color = backgroundcolor, fill = backgroundcolor!20] (7.7,1.6) rectangle (10.7,0.3);
\node [color = backgroundcolor] at (9.2,1.2) {impedance};
\node [color =backgroundcolor] at (9.2,0.7) {identification};

\node [color=backgroundcolor] at (5.3,0.6) {excitation};

\draw[scale=0.5,color=backgroundcolor] (11.1,2.2) -- (11.2,3.2) -- (11.2,2.4) -- (11.3,2.8) -- (11.4,2) -- (11.4,2.9) -- (11.5,2.3) -- (11.6,3.1) -- (11.6,2.1) -- (11.7,2.9) -- (11.7,2.3) -- (11.7,2.8) -- (11.8,2.5) -- (11.9,2.9) -- (11.9,2.4) -- (12,3.1) -- (12,2.1) -- (12.1,3.1) -- (12.1,2.7) -- (12.2,2.4) -- (12.3,2.8) -- (12.3,2.5) -- (12.4,2.7) -- (12.4,2.5) -- (12.5,2.8) -- (12.6,2.3) -- (12.6,2.9) -- (12.7,2.4) -- (12.8,2.8) -- (12.8,2.5) -- (12.8,2.6) -- (12.9,3.1) -- (13,2.2) -- (13,2.7) -- (13.1,2.5);

\draw[fill=backgroundcolor!40,color=backgroundcolor!40,opacity=0.5] (11.4,1.1) node (v2) {} -- (11.4,3.5) -- (13,3.5) -- (13,3.8) -- (13.5,3.35) -- (13,2.9) -- (13,3.2) -- (11.7,3.2) -- (11.7,1.1) -- (11.4,1.1);
\node [backgroundcolor] at (8.8,2.2) {$i$};
\node [backgroundcolor] at (9.9,2.2) {$v$};

\draw (10.7,2.8) -- (10.7,2.4);
\node at (10.7,3.15) {PCC};
\draw [backgroundcolor,-latex](5.3,0.9) -- (5.3,1.9);
\end{tikzpicture}
}
    \vspace{-0.8cm}
    \caption{Sketch of the proposed grid impedance identification setup.}
     \vspace{-0.3cm}
    \label{fig:scheme}
\end{figure}
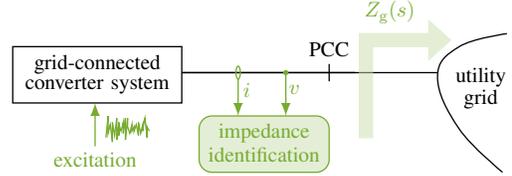
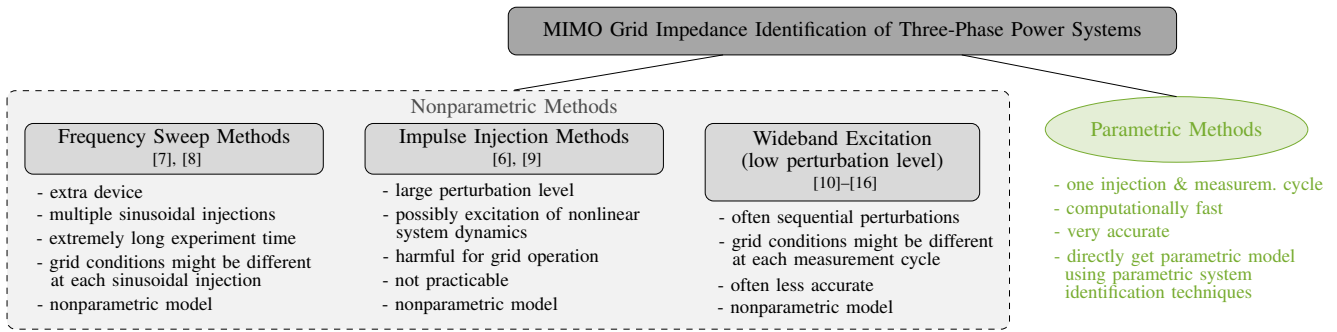
\begin{figure*}[t]
    \centering
    \resizebox {1\textwidth} {!} {
\begin{tikzpicture}
\draw [fill = black!5,rounded corners = 4, dashed] (-13.9,3.6) rectangle (2.9,-0.4);
\draw [rounded corners = 4,fill = black!35] (-5.5,5) rectangle (5.7,4.2);
\draw   [rounded corners = 4,fill = black!15](-13.6,3.05) rectangle (-8.6,2.2);

\draw  [rounded corners = 4,fill = black!15] (-7.9,3.05) rectangle (-2.9,2.2);

\draw  [rounded corners = 4,fill = black!15] (-2.2,3.05) rectangle (2.4,1.75);
\draw [color=backgroundcolor,fill= backgroundcolor!20] (5.7,2.95) ellipse (2.2 and 0.55);

\node [scale = 1] at (0.1,4.6) {{MIMO Grid Impedance Identification of Three-Phase Power Systems}};

\node at (-11.1,2.8) {Frequency Sweep Methods};

\node [scale = 0.8] at (-11.1,2.45) {\cite{francis2011algorithm,huang2009small}};
\node at (-5.4,2.8) {Impulse Injection Methods};

\node [scale = 0.8] at (-5.4,2.45) {\cite{cespedes2012online,liu2020analysis}};

\node at (0.1,2.8) {Wideband Excitation};
\node [color=backgroundcolor] at (5.7,2.95) {Parametric Methods};

\node [scale = 0.8] at (0.1,2.05) {\cite{xiao2007novel,martin2013wide,riccobono2017noninvasive,luhtala2018implementation,roinila2017mimo,shen2013three,xiao2022rapid}};

\node [scale = 0.9] at (-12.51,1.9) {- extra device};
\node [scale = 0.9 ] at (-11.23,1.1) {- extremely long experiment time};

\node [scale = 0.9 ] at (0.33,1.05) {- grid conditions might be different};
\node[scale = 0.9 ] at (0,0.75) {at each measurement cycle};

\node [scale = 0.9 ] at (-11.1,0.7) {- grid conditions might be different};
\node[scale = 0.9 ] at (-11.4,0.4) {at each sinusoidal injection};
\node[scale = 0.9 ] at (-11.4,1.5) {- multiple sinusoidal injections};

\node [scale = 0.9] at (-6,1.9) {- large perturbation level};
\node [scale = 0.9] at (-5.79,0.8) {- harmful for grid operation};

\node[scale = 0.9] at (-6.55,0.4) {-  not practicable};
\node [scale = 0.9] at (-5.45,1.5) {- possibly excitation of nonlinear};
\node [scale = 0.9] at (-6.28,1.2) {system dynamics};

\node [scale = 0.9] at (-0.68,0.35) {- often less accurate};
\node  [scale = 0.9] at (0.05,1.45) {- often sequential perturbations};

\node  [scale = 0.9,backgroundcolor] at (5.9,2) {- one injection \& measurem. cycle};

\node  [scale = 0.9,backgroundcolor] at (5.06,1.6) {- computationally fast};
\node  [scale = 0.9,backgroundcolor] at (4.62,1.2) {- very accurate};
\node [scale = 0.9,backgroundcolor] at (5.67,0.8) {- directly get parametric model};

\node [scale= 0.9] at (-6.14,0) {- nonparametric model};
\node [scale= 0.9] at (-0.52,-0.05) {- nonparametric model};
\node at (0.1,2.4) {(low perturbation level)};
\draw (-1.9,4.2) -- (-5.4,3.61); 
\draw (2.1,4.2) -- (5.7,3.5); 

\node [scale= 0.9] at (-11.925,0) {- nonparametric model};

\node [color=black!70] at (-5.4,3.3) {Nonparametric Methods};

\node [backgroundcolor,scale=0.9] at (5.4,0.5) {using parametric system};
\node [backgroundcolor,scale=0.9] at (5.4,0.2) {identification techniques};

\end{tikzpicture}
}
    \vspace{-0.8cm}
    \caption{Overview of existing approaches for MIMO grid impedance identification/measurement of three-phase power systems.}
    \label{fig:mindmap}
    \vspace{-0.3cm}
\end{figure*}

Other grid impedance identification methods use wideband signals to excite all frequencies of interest simultaneously, thereby significantly reducing the measurement time. One class of wideband excitation methods with high peak injection power are impulse-injection based methods \cite{cespedes2012online,liu2020analysis}. However, since their perturbation level is quite large, they may excite the nonlinear response of the grid, and additionally be harmful to the grid operation. Therefore, they do not work very well in practice and are typically not applied.

More advanced wideband excitation methods with smaller perturbation level use band-limited white noise \cite{xiao2007novel}, pseudo random binary sequences (PRBS) \cite{martin2013wide,riccobono2017noninvasive,luhtala2018implementation,roinila2017mimo}, chirp signals \cite{shen2013three}, quadratic-amplitude modulation multitone signals \cite{xiao2022rapid}, etc. However, most of these methods rely on sequential perturbations of each phase \cite{martin2013wide} or $dq$-axis component \cite{riccobono2017noninvasive} (thereby neglecting cross-coupling effects of the MIMO grid impedance), or apply multiple linear independent perturbations \cite{shen2013three,xiao2022rapid,xiao2007novel}, such that several measurement cycles are required. This is not only time-consuming, but also vulnerable to variable grid conditions between the perturbation cycles, resulting in a lower identification accuracy. 

Finally, once the identification data is obtained, all prevalent methods for grid impedance identification \cite{francis2011algorithm,huang2009small,cespedes2012online,liu2020analysis,xiao2007novel,martin2013wide,riccobono2017noninvasive,luhtala2018implementation,roinila2017mimo,shen2013three,xiao2022rapid} typically apply the fast Fourier transform to extract harmonic voltage and current responses at each frequency, and pointwise compute the grid impedance curve in the frequency domain. However, such a \textit{nonparametric} grid impedance model is generally cumbersome and cannot be immediately used for stability analysis or control design of converters.

To this end, we propose a \textit{parametric} approach to directly identify a parametric MIMO system model of the power grid in the $dq$ frame. It relies on injecting wideband excitation signals in the converter’s controller and allows to identify the grid impedance in closed loop during one injection and measurement cycle. In this regard, our approach is computationally fast and thus highly accurate even during variable grid conditions. Furthermore, the grid impedance model can be directly obtained from collected time-domain current and voltage data by applying parametric system identification techniques, such that point-wise Fourier computations are avoided. This makes the use of the identified model significantly easier and allows for an immediate stability analysis of the power system or an adaptation of the converter controller. Depending on the desired model structure to be identified, our approach is compatible
with any parametric system identification method, e.g., prediction error or subspace methods \cite{ljung1987theory}. Moreover, since the excitation signals have a small perturbation level,
our approach does not deteriorate the ongoing grid operation
and is well-suited for online grid impedance identification during real-time operation. 

 Finally, although parametric system identification techniques are in theory well-developed and find use in many applications, e.g., also to identify converter impedances \cite{gong2020dq}, their utilization in a grid impedance identification setup is vastly underexplored in power systems applications. Namely, their implementation is challenging and requires special care, e.g., when it comes to excitation power or model order selection. However, we will demonstrate in numerical experiments how the proposed parametric approach performs favorably (in some metrics highly superior) compared to prevalent nonparametric methods. 

The remainder of this paper is structured as follows. In \cref{sec:grid_impedance_identification}, we provide a tutorial introduction to dynamic grid impedance identification, where we present a general grid-connected converter configuration used for wideband excitation and data collection in closed loop. \cref{sec:sysID} gives an overview of the grid impedance identification procedure based on conventional parametric system identification techniques. In \cref{sec:numerical_experiments}, we demonstrate the effectiveness of our parametric approach in numerical case studies and compare it with prevalent nonparametric methods for grid impedance identification. \cref{sec:conclusion} concludes the paper.

\section{Excitation and Measurement Configuration}\label{sec:grid_impedance_identification}
This section presents the grid-connected converter configuration used for wideband excitation and data collection, and gives a tutorial introduction to grid impedance identification.

\subsection{Small-Signal Grid Impedance}
Consider the three-phase power converter system in \cref{fig:scheme} which is connected to the utility grid at the PCC. We aim to identify the dynamic small-signal grid impedance characterized by a parametric MIMO transfer function $Z_\mathrm{g}(s)$, which relates the terminal voltages and currents of the power grid at the PCC. In particular, the small-signal grid impedance describes the linearized power grid dynamics at the current steady-state operating point. 

By applying the $dq$-coordinate transformation (Park transformation), a three-phase power grid under balanced conditions can be represented as a $dq$-coordinate system with a constant steady-state operating point, where the steady-state grid frequency $\omega_\mathrm{g}$ is considered as the rotating frequency of the $dq$ frame. Consequently, the ac voltages and currents at the PCC in $abc$ coordinates become dc quantities in $dq$ coordinates, i.e., admitting a constant steady-state. Based on this, the dynamic small-signal impedance $Z_\mathrm{g}(s)$ of the power grid can be modeled with four transfer functions $Z_\mathrm{g,dd}(s),\,Z_\mathrm{g,dq}(s),\,Z_\mathrm{g,qd}(s)$ and $Z_\mathrm{g,qq}(s)$, respectively, such that the small-signal voltages $\Delta v_\mathrm{dq}(s)$ and currents $\Delta i_\mathrm{dq}(s)$ at the PCC (deviating from the respective steady-state value) are related as
\begin{align}\label{eq:Z_g}
    \underset{=\Delta v_\mathrm{dq}(s)}{\underbrace{\begin{bmatrix}
    \Delta v_\mathrm{d}(s)\\ \Delta v_\mathrm{q}(s) 
    \end{bmatrix}}} =
    \underset{=Z_\mathrm{g}(s)}{\underbrace{\begin{bmatrix}
        Z_\mathrm{g,dd}(s)& Z_\mathrm{g,dq}(s)\\ Z_\mathrm{g,qd}(s)& Z_\mathrm{g,qq}(s)
    \end{bmatrix}}}
    \underset{=\Delta i_\mathrm{dq}(s)}{\underbrace{\begin{bmatrix}
    \Delta i_\mathrm{d}(s)\\ \Delta i_\mathrm{q}(s) 
    \end{bmatrix}}}.
\end{align}

\subsection{Grid-Connected Converter System}
\cref{fig:inverter_scheme} depicts a three-phase power converter system which is used for wideband excitation and grid impedance identification. It is based on a three-phase voltage source converter (VSC) with IGBT switches, and connected to the power grid at the PCC via an $LCL$ filter. In the control scheme of the converter, the three-phase voltage and current signals are also represented in a $DQ$ synchronous reference frame, where, however, the rotating frequency of the $DQ$ frame is generated by a phase-locked-loop (PLL), which tracks the frequency at the $LCL$ capacitor \cite{chung2000phase}. Notice that we use capital letters to denote the local PLL-based $DQ$ frame of the converter control in order to differentiate from the global $dq$ frame of the small-signal grid impedance in \cref{eq:Z_g} which relies on the steady-state grid frequency $\omega_\mathrm{g}\hspace{-0.5mm}=\hspace{-0.5mm}\dot{\theta}_\mathrm{g}$. The converter system is based on a standard control scheme in the PLL's $DQ$ frame \cite{yazdani2010voltage}, using PI controllers inside the current control loop to make the converter-side current ($i_\mathrm{c,D}$ and $i_\mathrm{c,Q}$) track their references ($i_\mathrm{c,D}^\mathrm{ref}$ and $i_\mathrm{c,Q}^\mathrm{ref}$). Finally, the converter voltage is generated by the pulse-width modulation (PWM) of the VSC. 

\subsection{Wideband Excitation \& Data Collection}
To extract the grid impedance $Z_\mathrm{g}(s)$, we inject uncorrelated wideband excitation signals with small perturbation levels (e.g., PRBS, random binary sequences (RBS), chirp signals, etc.) in the converter's control loop to excite the power grid during online operation (\cref{fig:inverter_scheme}). The resulting closed-loop voltage and current responses at the PCC are then measured and collected in the form of discrete-time samples $v_\mathrm{abc}(k)$ and $i_\mathrm{abc}(k)$ with sampling index $k$, which, after $dq$ transformation and offset removal, will be later used to calculate the small-signal grid impedance (see \cref{sec:sysID}).

\begin{figure}[t]
    \centering
    \usetikzlibrary{circuits.ee.IEC}
\usetikzlibrary{arrows}
\tikzstyle{roundnode} =[circle, draw=blue!60, fill=blue!5, scale = 0.5]

\resizebox {0.5\textwidth} {!} {

\begin{tikzpicture}[circuit ee IEC,scale=0.47, every node/.style={scale=0.65}]

\draw (2.5,3.6) -- (3.5,3.6);
\draw (2.5,1.6) -- (3.5,1.6);
\draw  (3.5,3.8) node (v13) {} rectangle (5.3,1.4);

\draw (4.6,3.5) -- (4.6,3.1) -- (4.1,2.85) node (v3) {};
\draw[-latex](4.1,2.35) -- (4.6,2.1) ;
\draw (4.6,2.1) -- (4.6,1.7);
\draw (4.1,3.1) -- (4.1,2.1);
\draw (4,3.1)--(4,2.1);
\draw (3.7,2.6) -- (4,2.6);

\draw (6,2.6) node (v7) {} to [inductor={yscale=1.5}] (8,2.6) node (v8) {};
\draw (8.9,2.6) node (v7) {} to [inductor={yscale=1.5}] (10.9,2.6) node (v8) {};

\draw (5.3,2.6) -- (6.1,2.6) node (v1) {}; 

\draw (10.6,2.6) -- (13,2.6) -- (14.4,2.6);

\node at (2.7,3.3) {$+$};
\node at (2.7,1.8) {$-$};
\node at (2.8,2.6) {$v_\mathrm{dc}$};

\node at (7,3.2) {$L_\mathrm{f,1}$};

\draw(2.45,3.6) circle (0.6 mm); 

\draw(2.45,1.6) circle (0.6 mm);

\draw[-latex] (4.4,0.6) -- (4.4,1.4);
\draw[rounded corners=2,fill=black!10]  (3.5,0.6) rectangle (5.3,-0.2);
\node at (4.4,0.2) {PWM};
\node at (4.4,4.15) {VSC};
\draw [-latex](4.4,-1.4) -- (4.4,-0.2);
\draw  (3.7,-1.4) rectangle (5.1,-2.3);
\draw (3.7,-2.3) -- (5.1,-1.4);
\node [scale= 0.9] at (4.1,-1.65) {abc};
\node [scale = 0.7] at (4.7,-2.05) {DQ};
\node at (3.7,-0.9) {$u_\mathrm{c,abc}^\mathrm{ref}$};
\draw[-latex] (4,-4) -- (4,-2.3); 
\draw[-latex] (4.8,-3.3) -- (4.8,-2.3);

\draw (4.8,-3.45) circle(1.5mm);

\draw (4,-4.15) circle(1.5mm);

\draw[-latex] (6.2,-3.45) -- (4.95,-3.45);
\draw [-latex] (6.2,-4.15)--(4.15,-4.15) ;
\draw[-latex,backgroundcolor] (4,-4.9) -- (4,-4.3); 
\draw [-latex,backgroundcolor](4.8,-4.9) -- (4.8,-3.6);
\node [backgroundcolor] at (4.4,-5.2) {wideband excitation};
\node [backgroundcolor] at (4.4,-5.7) {signal injection};
\draw [rounded corners = 3,fill=black!10] (6.2,-3.1) rectangle (9.9,-4.5);
\draw (8.4,2.6) -- (8.4,1.95);
\fill (8.4,2.6) circle(0.6mm);
\draw (8.1,1.95) -- (8.7,1.95); 
\draw (8.1,1.8) -- (8.7,1.8); 
\draw (8.4,1.8) -- (8.4,1.2);
\draw (8.25,1.2) -- (8.55,1.2);
\draw (8.35,1.15) -- (8.45,1.15);
\node at (7.7,1.85) {$C_\mathrm{f}$};

\draw[color=black]  (5.6,2.6) ellipse (0.05 and 0.15);

\draw  (6.2,-1.4) rectangle (7.6,-2.3);
\draw[-latex,color=black] (5.6,2.45) --(5.6,-1.9)-- (6.2,-1.9);
\draw (6.2,-2.3) -- (7.6,-1.4);
\draw [-latex] (3,-1.85) -- (3.7,-1.85);
\node at (2.55,-1.85) {$\theta_\mathrm{pll}$};
\node [scale=0.9] at (6.6,-1.65) {abc};
\node [scale=0.7] at (7.2,-2.05) {DQ};

\node[color=black] at (6.25,-0.7) {$i_\mathrm{c,abc}$};
\draw [-latex,color=black] (8.9,2.6) -- (8.9,-1.4);

\fill [color=black](8.9,2.6) circle(0.6mm);
\draw [rounded corners = 2,fill=black!10] (8.5,-1.4) rectangle (9.9,-2.3);
\node at (9.2,-1.85) {PLL};
\draw[-latex] (6.4,-2.3) -- (6.4,-3.1); 
\draw [-latex](7.4,-2.3) -- (7.4,-3.1);
\node at (6.9,-2.75) {$i_\mathrm{c,D}$};
\node at (7.9,-2.75) {$i_\mathrm{c,Q}$};
\draw [-latex](8.5,-1.85) -- (7.6,-1.85);
\node at (8.1,-1.5) {$\theta_\mathrm{pll}$};
\node at (8.1,-3.55) {$DQ$-current};
\node [color=black] at (8.2,-0.7) {$v_\mathrm{c,abc}$};
\node at (8.1,-4.05) {control};
\draw[-latex] (10.6,-3.45) -- (9.9,-3.45);
\draw [-latex](10.6,-4.15) -- (9.9,-4.15);
\node at (11.1,-3.45) {$i_\mathrm{c,D}^\mathrm{ref}$};
\node at (3.5,-2.75) {$u_\mathrm{c,Q}^\mathrm{ref}$};
\node at (5.35,-2.75) {$u_\mathrm{c,D}^\mathrm{ref}$};
\node at (11.1,-4.15) {$i_\mathrm{c,Q}^\mathrm{ref}$};

\draw [color=backgroundcolor] (12.5,2.6) ellipse (0.05 and 0.15);
\draw[-latex,color=backgroundcolor] (12.5,2.45) -- (12.5,2) -- (12.5,0.4);
\draw [-latex,color=backgroundcolor](11.5,2.6) -- (11.5,0.4);
\fill[color=backgroundcolor] (11.5,2.6) circle(0.6mm);
\draw [rounded corners = 3, color=backgroundcolor, fill = backgroundcolor!20] (10.5,-1.3) rectangle (13.5,-2.4);
\node [color=backgroundcolor] at (12,-1.65) {impedance};
\node  [color=backgroundcolor] at (12,-2.05) {identification};
\draw[fill=backgroundcolor!40,color=backgroundcolor!40,opacity=0.5] (13.5,1.6) node (v2) {} -- (13.5,3.8) -- (15.3,3.8) -- (15.3,4.1) -- (15.8,3.65) -- (15.3,3.2) -- (15.3,3.5) -- (13.8,3.5) -- (13.8,1.6) -- (13.5,1.6);
\node [color=backgroundcolor] at (14.55,4.15) {$Z_\mathrm{g}(s)$};

\node[backgroundcolor] at (12.05,2.2) {$i_\mathrm{abc}$};

\draw  (14.4,2.8) rectangle (15.5,2.4);
\draw (15.5,2.6) -- (16.2,2.6) -- (16.2,2.2);
\draw  (16.2,1.9) ellipse (0.3 and 0.3);
\draw (16.2,1.6) -- (16.2,1.2);
\draw (16.05,1.2) -- (16.35,1.2);
\draw (16.15,1.15) -- (16.25,1.15);
\draw  [dashed, rounded corners =3,color=backgroundcolor](14.1,3) rectangle (16.7,0.9);
\node [color=backgroundcolor] at (15.4,0.55) {three-phase};
\node [color=backgroundcolor] at (15.4,0.1) {grid};
\draw  plot[smooth, tension=.7] coordinates {(16,1.85) (16.1,2.05) (16.3,1.75) (16.4,1.95)};
\draw (13,2.8) -- (13,2.4);
\node at (12.85,3.15) {PCC};

\draw (4.6,1.9) -- (4.9,1.9) -- (4.9,2.5); 
\draw (4.9,2.8) -- (4.9,3.3) -- (4.6,3.3); 
\draw (4.75,2.8) -- (5.05,2.8); 
\draw (4.9,2.8) -- (4.7,2.5) -- (5.1,2.5) -- (4.9,2.8);

\draw (7.8,2.6) -- (8.9,2.6);
\node at (9.9,3.2) {$L_\mathrm{f,2}$};

\node [backgroundcolor,scale = 0.7] at (10.2,1.1) {measurement};
\node [backgroundcolor,scale= 0.7] at (10.8,0.8) {noise};

\draw[-latex,scale=0.5,backgroundcolor] (23.7,3.6) -- (24.1,2.8) -- (24.2,3.4) -- (24.8,2.1);
\draw[-latex,scale=0.5,backgroundcolor] (21.7,3.6) -- (22.1,2.8) -- (22.2,3.4) -- (22.8,2.1);
\node [backgroundcolor] at (11,2.2) {$v_\mathrm{abc}$};
\draw  [color=backgroundcolor](11.3,0.4) rectangle (12.7,-0.5);
\draw [color=backgroundcolor](11.3,-0.5) -- (12.7,0.4);
\node  [scale=0.9,backgroundcolor] at (11.7,0.15) {abc};
\node  [scale=0.9, backgroundcolor] at (12.3,-0.2) {dq};
\draw [-latex,backgroundcolor](11.5,-0.5) -- (11.5,-1.3); 
\draw [-latex,backgroundcolor](12.5,-0.5) -- (12.5,-1.3);
\node [backgroundcolor]at (11,-0.8) {$v_\mathrm{dq}$};
\node [backgroundcolor]at (12.1,-0.8) {$i_\mathrm{dq}$};

\draw [-latex,backgroundcolor](10.6,-0.05) -- (11.3,-0.05);
\node [backgroundcolor]at (10.3,-0.05) {$\theta_\mathrm{g}$};
\end{tikzpicture}

}
    \vspace{-0.8cm}
    \caption{One-line diagram of three-phase grid-connected power converter system used for wideband excitation and grid impedance identification. The proposed approach can be also applied to other type of converter controls.}
    \vspace{-0.3cm}
    \label{fig:inverter_scheme}
\end{figure}
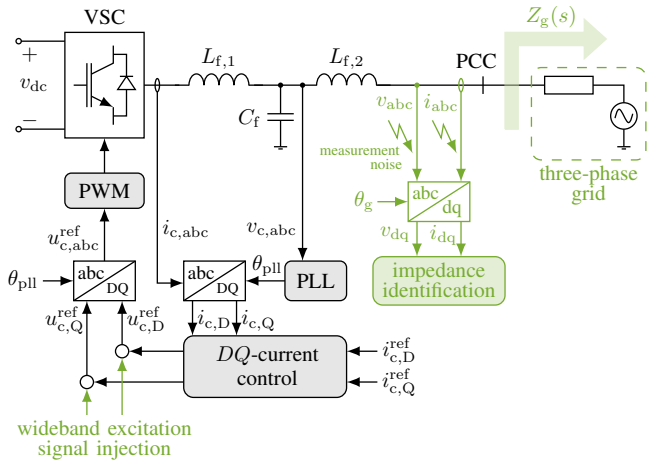
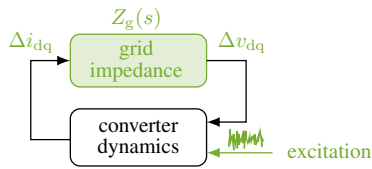
\begin{figure}[t]
    \centering
    \usetikzlibrary{circuits.ee.IEC}
\usetikzlibrary{arrows}

\resizebox {0.38\textwidth} {!} {
\tikzstyle{roundnode} =[circle, draw=blue!60, fill=blue!5, scale = 0.5]
\begin{tikzpicture}[circuit ee IEC,scale=0.47, every node/.style={scale=0.65}]

\draw [rounded corners = 3,backgroundcolor,fill=backgroundcolor!20] (3.8,4.9) rectangle (6.9,3.7);

\node [color=backgroundcolor] at (5.3,5.25) {$Z_\mathrm{g}(s)$};

\draw [rounded corners = 3] (6.9,3.2) rectangle (3.8,1.9);
\node at (5.35,2.8) {converter};
\node at (5.35,2.3) {dynamics};

\node [color=backgroundcolor] at (9.7,2.25) {excitation};

\draw[scale=0.4,color=backgroundcolor] (18.5,5.85) -- (18.6,6.85) -- (18.6,6.05) -- (18.7,6.45) -- (18.8,5.65) -- (18.8,6.55) -- (18.9,5.95) -- (19,6.75) -- (19,5.75) -- (19.1,6.55) -- (19.1,5.95) -- (19.1,6.45) -- (19.2,6.15) -- (19.3,6.55) -- (19.3,6.05) -- (19.4,6.75) -- (19.4,5.75) -- (19.5,6.75) -- (19.5,6.35) -- (19.6,6.05) -- (19.7,6.45) -- (19.7,6.15) -- (19.8,6.35) -- (19.8,6.15) -- (19.9,6.45) -- (20,5.95) -- (20,6.55) -- (20.1,6.05) -- (20.2,6.45) -- (20.2,6.15) -- (20.2,6.25) -- (20.3,6.75) -- (20.4,5.85) -- (20.4,6.35) -- (20.5,6.15);

\draw [backgroundcolor,-latex](8.4,2.2) -- (6.9,2.2);
\draw[-latex] (3.8,2.5) -- (2.9,2.5) -- (2.9,4.3) -- (3.8,4.3); 

\draw [-latex](6.9,4.3) -- (7.8,4.3) -- (7.8,2.9) -- (6.9,2.9);
\node [backgroundcolor] at (5.3,4.55) {grid};
\node [backgroundcolor] at (2.9,4.7) {$\Delta i_\mathrm{dq}$};
\node [backgroundcolor] at (7.7,4.7) {$\Delta v_\mathrm{dq}$};
\node [backgroundcolor]at (5.3,4.05) {impedance};
\node at (0.2,3.1) {};
\end{tikzpicture}
}
    \vspace{-0.8cm}
    \caption{Small-signal block diagram of grid impedance identification problem.}
    \vspace{-0.3cm}
    \label{fig:block_diagram}
\end{figure}

\cref{fig:block_diagram} shows a block diagram of the resulting \textit{closed-loop} grid impedance identification problem, where the input/output perspective of the to-be-identified grid impedance results from the associated electrical circuit equations. Notice that the measured closed-loop voltage and current samples at the PCC used for system identification are affected by the converter dynamics, while being additionally noisy due to the IGBT switching harmonics of the VSC and measurement noise. Therefore, from a system identification point of view, the configuration is rather special, and it is challenging to extract an accurate model of the \textit{continuous-domain} grid impedance $Z_\mathrm{g}(s)$ at the PCC from such noisy and distorted \textit{discrete-time} samples. To tackle this problem, it is therefore essential to choose adequate amplitudes of the injected excitation signals, such that they can provide a good signal-to-noise ratio at the PCC, while at the same time being low enough to avoid effects of nonlinear dynamic phenomena or a harmful impact on the grid operation. Furthermore, as will be discussed in \cref{sec:model_order_structure}, a higher model order selection of the identified discrete-domain model (compared to the true system order of the continuous-domain grid impedance) has been observed to compensate for discretization inaccuracies, and, in the face of the noise, additionally provides more flexibility to the model structure. 

\section{Grid Impedance Identification}\label{sec:sysID}
Once the $v_\mathrm{abc}(k)$ and $i_\mathrm{abc}(k)$ measurements are obtained, the grid impedance identification can be performed. The procedure consists of four steps, which are described in \cref{sec:signal_preprocessing,sec:model_order_structure,sec:algorithm,sec:post_processing} in the following:

\subsection{Signal Preprocessing}\label{sec:signal_preprocessing}
\subsubsection*{dq-transformation} To identify the grid impedance $Z_\mathrm{g}(s)$ as in \cref{eq:Z_g}, the $v_\mathrm{abc}(k)$ and $i_\mathrm{abc}(k)$ measurements are transformed into the global $dq$ frame, where the rotating frequency of this $dq$ frame is a \textit{constant} approximation of the detected grid frequency $\omega_\mathrm{g}$ at the current steady-state operating point (\cref{fig:inverter_scheme}). This constant frequency value can be obtained by, for instance, wide-area measurements, state-estimation techniques, or an averaged PLL-measurement. One can alternatively generate the rotating frequency of the grid impedance's $dq$ frame by the converter's PLL. This is, however, generally less accurate, as the \textit{nonconstant} PLL frequency $\omega_\mathrm{pll}=\dot{\theta}_\mathrm{pll}$ undergoes some dynamic response due to the presence of noise and excitation.

\subsubsection*{Offset removal} Next, the measured steady-state values of the transformed $v_\mathrm{dq}(k)$ and $i_\mathrm{dq}(k)$ signals are subtracted to obtain the small-signal quantities $\Delta v_\mathrm{dq}(k)$ and $\Delta i_\mathrm{dq}(k)$.

\subsubsection*{Prefiltering} Finally, prefiltering both $\Delta v_\mathrm{dq}(k)$ and $\Delta i_\mathrm{dq}(k)$ can be (optionally) considered. Prefiltering emphasizes or de-emphasizes certain frequency content, and therefore acts as frequency weighting (for details see \cite{ljung1987theory}).

\subsection{Model Structure \& Order Selection}\label{sec:model_order_structure}
To identify a parametric grid impedance model, the desired model structure has to be selected. It is noted that the grid impedance model will be identified in the discrete domain based on discrete-time current and voltage data, which can later be converted to the continuous domain, if required (see \cref{sec:post_processing}). In the following, let $u(k):=\Delta i_\mathrm{dq}(k)\in \mathbb{R}^p$ (with $p=2$) and $y(k):=\Delta v_\mathrm{dq}(k)\in \mathbb{R}^m$ (with $m=2$) denote the system input and output, respectively. 

Typical parametric model structures are transfer function and state-space models. The former are described by rational functions with numerator and denominator coefficients, i.e.,  
\begin{align}\label{eq:tf}
    y(k) &= G(q,\psi)u(k) + H(q,\psi)e(k)
\end{align}
where $e(k)\in\mathbb{R}^m$ is a white noise disturbance, $\psi\in\mathbb{R}^d$ is the parameter vector of numerator and denominator coefficients to be identified, and $q$ is a shift operator meaning $q^{-1}x(k)=x(k-1)$. $G(q,\psi)$ and $H(q,\psi)$ are the $m\times p$ input transfer matrix and the $m\times m$ noise transfer matrix, respectively, whose entries are rational functions of $q^{-1}$. Depending on the parametrization of $G(q,\psi)$ and $H(q,\psi)$ in \cref{eq:tf}, different model structures can be obtained. The probably simplest structure to be identified is the so-called \textit{ARX} model \cite{ljung1987theory}
\begin{align}\label{eq:arx}
\begin{split}
    \hspace{-0.5mm}G(q,\psi)\hspace{-0.5mm}=\hspace{-0.5mm}A^{-1}(q,\psi)B(q,\psi),\quad H(q,\psi)\hspace{-0.5mm}=\hspace{-0.5mm}A^{-1}(q,\psi),
\end{split}
\end{align}
where \textit{AR} refers to the \textit{autoregressive} part $A(q,\psi)y(k)$ and \textit{X} to the \textit{exogenous} input $B(q,\psi)u(k)$. The matrix polynomials $A(q,\psi)$ and $B(q,\psi)$ are matrices of dimensions $m\hspace{-0.5mm}\times\hspace{-0.5mm}m$ and $m\hspace{-0.5mm}\times\hspace{-0.5mm}p$, respectively, whose entries are polynomials in $q^{-1}\hspace{-1mm}$, i.e., 
\begin{align}
\begin{split}
    A(q,\psi) & = I + A_1 q^{-1} + A_2 q^{-2} + ... + A_{n_\mathrm{a}} q^{-n_\mathrm{a}}\\
    B(q,\psi) & = B_1 q^{-1} + B_2 q^{-2} + ... + B_{n_\mathrm{b}} q^{-n_\mathrm{b}},
\end{split}
\end{align}
with parameter matrices $A_i$ and $B_i$ to be identified. There also exist other parametrizations of $G(q,\psi)$ and $H(q,\psi)$ in \cref{eq:tf}, defining other model and noise disturbance structures, e.g., so-called ARMAX, OE or BJ structures. The identification task, however, is more complex in these cases, and hence not considered in this paper (for details see \cite{ljung1987theory}).

On the other hand, discrete-time state-space models relate the input, noise, and output signals via first-order difference equations 
using an auxiliary state vector $x(k)\hspace{-0.5mm}\in\hspace{-0.5mm}\mathbb{R}^n$, i.e., 
\begin{align}\label{eq:ss}
\begin{split}
    x(k+1) &= A x(k) + B u (k) + w(k)\\
    y(k) &= C x(k) + D u(k) + \nu(k), 
    \end{split}
\end{align}
where $w(k)\hspace{-0.5mm}\in\hspace{-0.5mm}\mathbb{R}^n$ and $\nu(k)\hspace{-0.5mm}\in\hspace{-0.5mm}\mathbb{R}^m$ are process and measurement noise disturbances, respectively (assumed to be white), and $A, B, C$ and $D$ the state-space matrices to be identified.

Notice that, for grid impedance identification, it is generally difficult to decide which model and disturbance structure is more reasonable (if not specified by the application configuration). However, we will later see in the numerical case studies in \cref{sec:numerical_experiments} that often several model structures can be effective. Moreover, it can be observed that a higher model order selection of the identified discrete-domain model (compared to the true system order of the continuous-domain grid impedance) allows for more flexibility in the model and disturbance structure in case of noise in a practical system identification setup, and additionally compensates for discretization inaccuracies. Therefore, it is often more accurate to first identify a higher-order model which achieves a close fit of the experimental data, and then to perform a model reduction, if sought \cite{hjalmarsson2005experiment,arnedo2008terminated}. 

\begin{remark} Unlike standard identification where sampling-hold inputs can be provided, the voltage and current signals in our setting are physical quantities measured from a continuous-time system, and one may encounter discretization inaccuracies.
To give some intuition on how to compensate for discretization inaccuracies with a higher model order selection of the identified discrete-domain model, recall the expression $e^{sT}=z$ with sampling period $T$ (or inversely $s=\tfrac{1}{T}\ln z$), which relates the equivalent characteristics in the $s$-domain to those in the $z$-domain \cite{oppenheim1999discrete}. Hence, if we consider the true continuous-domain grid impedance $Z_\mathrm{g}(s)$ in the $s$-domain, an exact discretization can be obtained as $Z_\mathrm{g}(s = \tfrac{1}{T}\ln z)$ (if it exists), which is not a rational transfer function. However, with the imposed discrete-time model structures in \eqref{eq:arx} and \eqref{eq:ss}, we obtain rational transfer functions after system identification, which can be easily used in system analysis and control synthesis. In respect thereof, if one was to approximate $Z_\mathrm{g}(s = \tfrac{1}{T}\ln z)$ by means of a rational transfer function expression in the $z$-domain, an immediate approach would be to resort to a finite length approximation of a series expansion of $\ln z$. In doing so, considering more terms in the series expansion of $\ln z$ can result in a better approximation accuracy of $\ln z$, which in turn is equal to selecting a higher order of the rational transfer function expression $Z_\mathrm{g}(z)$ in the $z$-domain.
\end{remark}
Note that due to the existence of noise in our practical system identification setup, a too high model-order selection of the identified discrete-domain model can lead to overfitting issues, which should be avoided. Hence, the choice of the model order turns out to be a trade-off between the compensation of model structure and discretization inaccuracies on the one hand, and overfitting issues on the other hand.\color{black}

Finally, a practical strategy to select an accurate model structure with an appropriate order, especially in case of an unknown system to be identified, would be to iteratively compare the fitting accuracy of different model structures and orders for an additional validation data set (see \cref{sec:post_processing}). It is then recommended to choose the model structure of the lowest order that provides a reasonably good fit, to reduce the model complexity and hence the computational requirements for subsequent analysis and control design applications. Alternatively, if one was to select an appropriate model structure and order without the need of additional validation data, one could iteratively evaluate some statistical criteria such as the Akaike Information Criterion (AIC) \cite{stoica2004model} for different model structures and orders.

\subsection{Parametric System Identification}\label{sec:algorithm}
Transfer function models \cref{eq:tf} as well as state-space models \cref{eq:ss} can be identified via parametric system identification techniques. The former are typically obtained via techniques based on \textit{prediction error methods (PEM)}, while the latter are usually computed via \textit{subspace methods} \cite{ljung1987theory}. Both of these methods have been widely discussed in literature, and there are commercial tools available to apply them in an easy way, e.g., the System Identification Toolbox
of Matlab \cite{ljung2022system}.

In the following, we will provide the key aspects of both methods to identify the associated parametric model. To do so, we consider the input-output data set of collected current and voltage samples over a time-interval $1\leq k\leq N_\mathrm{s}$.

\subsubsection{Prediction Error Methods (PEM)} 
The one-step ahead predictor of the transfer function model in \cref{eq:tf}, and thus of the ARX model in \cref{eq:arx}, is given by the conditional expectation of $y(k)$ over the past data (see \cite{ljung1987theory} for a derivation), i.e., 
\begin{align}
    \hat{y}(k|\psi):\hspace{-0.9mm}&= H^{-1}(q,\psi)G(q,\psi)u(k)+[I-H^{-1}(q,\psi)]y(k)\nonumber\\
    &= B(q,\psi)u(k)+[I-A(q,\psi)]y(k).\label{eq:predictor}
\end{align}

Next, by introducing the $(n_am\hspace{-0.3mm}+\hspace{-0.3mm}n_bp)$\,-\,dimensional vector\\ 
$\varphi(k)\hspace{-0.3mm}=\hspace{-0.3mm}[-y(k\hspace{-0.3mm}-\hspace{-0.3mm}1)^\top\hspace{-0.25mm}...\,-\hspace{-0.25mm}y(k\hspace{-0.3mm}-\hspace{-0.3mm}n_a)^\top\hspace{-0.25mm}u(k\hspace{-0.25mm}-\hspace{-0.25mm}1)^\top\hspace{-0.25mm}...\,u(k\hspace{-0.25mm}-\hspace{-0.25mm}n_b)^\top\hspace{-0.25mm}]^\top$\\
and the $(n_am\hspace{-0.3mm}+\hspace{-0.3mm}n_bp)\hspace{-0.9mm\times p}$ \,\,matrix $\Psi \hspace{-0.35mm}= \hspace{-0.35mm}[A_1 \,...\, A_{n_a} B_1 \,...\, B_{n_b}]$,\\
the one-step ahead predictor in \cref{eq:predictor} can be rewritten as a linear regression model $\hat{y}(k,\psi)=\Psi^\top\varphi(k)$, where $\varphi(k)$ is the regression vector and $\Psi$ the parameter matrix to be identified. To do so, a least-squares linear regression problem is solved to minimize the prediction error $\varepsilon(k,\psi)=y(k)-\Psi^\top\varphi(k)$. Notice that for other transfer function structures (e.g, ARMAX, OE, and BJ), pseudo-linear regression problems have to be solved via more advanced optimization methods (e.g., Gauss-Newton, etc. \cite{ljung1987theory}).

\subsubsection{Subspace Methods} 
Considering \cref{eq:ss}, we can find that
\begin{align}
    y(k+r) &= CA^rx(k)+CA^{r-1}Bu(k)+ ...\,+\nonumber\\
    &+ CBu(k+r-1)+Du(k+1)\,+\label{eq:ss_expanded}\\
    &+CA^{r-1}w(k)+ ... + Cw(k+r-1)+\nu(k+r),\nonumber
\end{align}
for $r\hspace{-0.25mm}\geq \hspace{-0.25mm}n\hspace{-0.25mm}+\hspace{-0.25mm}1$. Next, for $N\hspace{-0.25mm}\leq\hspace{-0.25mm} N_\mathrm{s}\hspace{-0.25mm}-\hspace{-0.25mm}r\hspace{-0.25mm}+\hspace{-0.25mm}1$, we define matrices 
\begin{align}
\begin{split}
    \mathbf{X}&=[x(1)\,...\,x(N)]\quad\mathbf{Y}=[Y_r(1)\,...\,Y_r(N)]\\
    \mathbf{U}&=[U_r(1)\,...\,U_r(N)] \quad\mathbf{V}=[V(1)\,...\,V(N)],
\end{split}
\end{align}
with the vectors $Y_r(k) = [y(k)^\top...\,\,y(k+r-1)^\top]^\top$ and $U_r(k)\hspace{-0.2mm}=\hspace{-0.2mm}[u(k)^\top...\,\,u(k\hspace{-0.1mm}+\hspace{-0.1mm}r\hspace{-0.2mm}-\hspace{-0.2mm}1)^\top]^\top\hspace{-2mm}$, and the $j$th block of $V$ as\\
$V_j(k) = CA^{j-2}w(k)+...+Cw(k+j-2)+\nu(k+j-2)$. Based on this, \cref{eq:ss_expanded} can be written as
\begin{align}\label{eq:ss_matrix}
    \mathbf{Y} = O_r \mathbf{X}+S_r\mathbf{U}+\mathbf{V},
\end{align}
where the matrix $S_r$ follows from inspection, and $O_r$ is the extended $mr\hspace{-0.5mm}\times\hspace{-0.5mm}n$ observability matrix of \cref{eq:ss}, i.e., 
\begin{align}\label{eq:observability_matrix}
   \textstyle O_r = \begin{bmatrix}
        C^\top&(CA)^\top& \dots & (CA^{r-1})^\top
    \end{bmatrix}^\top,
\end{align}
which is later used to obtain the $A$ and $C$ matrices in \cref{eq:ss}.

To extract the extended observability matrix $O_r$, both sides of \cref{eq:ss_matrix} can be correlated with quantities that eliminate the term $S_r\mathbf{U}$, and make the noise influence from $\mathbf{V}$ disappear as $N\rightarrow \infty$. In particular, we can rewrite \cref{eq:ss_matrix} as
\begin{align}\label{eq:cal_O}
\mathcal{O}:=\mathbf{Y}\mathcal{T}=O_r\mathbf{X}\mathcal{T}+\mathcal{E},
\end{align}
where the $N\hspace{-0.5mm}\times\hspace{-0.5mm}s$ matrix $\mathcal{T}$ with $s\geq n$ is solely made up of input-output data, and encodes the elimination of $S_r\mathbf{U}$ and $\mathbf{V}$ in \cref{eq:ss_matrix} via orthogonal projection and instrumental variable techniques \cite{ljung1987theory}. $\mathcal{E}$ is a vanishing noise matrix for $N\rightarrow\infty$. 

Given $\mathcal{O}$, the matrix $O_r$ can be recovered by employing singular value decomposition techniques. The $\hat{A}$ and $\hat{C}$ matrices in \cref{eq:ss} are then obtained by using the first block row of $O_r$ and applying the shift property. Finally, the $\hat{B}$ and $\hat{D}$ matrices are estimated from a linear least-squares problem using the predictor $\hat{y}(k|B,D)\hspace{-0.4mm}=\hspace{-0.4mm}\hat{C}(qI\hspace{-0.4mm}-\hspace{-0.4mm}\hat{A})^{-1}\hspace{-0.35mm}Bu(k)\hspace{-0.4mm}-\hspace{-0.4mm}Du(k)$.

Notice that PEM are typically simpler than subspace methods and can be computed very quickly, therefore making them possibly the preferred choice for real-time applications such as online grid impedance identification. 

\subsection{Model Validation \& Post-Processing} \label{sec:post_processing}
After identifying an unknown system, the identified model can be validated by comparing the model response with the measured response under the same input. In terms of model transformations, once the ARX model has been identified, we translate the difference equation obtained (see \cref{eq:tf}) into a discrete transfer function in the $z$-domain. In addition, if sought, both the identified discrete transfer function and state-space model can be converted to the continuous domain, and/or model order reduction techniques can be applied to further reduce the computational effort in subsequent applications.

\renewcommand{\arraystretch}{1.2}
\begin{table}[b!]\scriptsize
    \centering
           \caption{Electrical Parameters of the Numerical Experiment}
           \vspace{-0.2cm}
               \begin{tabular}{c||c|c}
     \toprule
         Parameter & Symbol & Value  \\ \hline
         $\hspace{-1mm}$Voltage, power \& freq. base$\hspace{-1mm}$& $V_\mathrm{b},\,S_\mathrm{b},\,f_\mathrm{b}$ & 380 V, 1.5 kVA, 50 Hz\\
         LCL filter components& $\hspace{-1mm}$$L_\mathrm{f,1},\,L_\mathrm{f,2},\,C_\mathrm{f}$$\hspace{-1mm}$& 0.08 p.u.\, 0.05 p.u., 0.08 p.u.\\
         Load component& $R_1$ & 2 p.u.\\
         Line 1 components& $R_2,\,L_2,\,C_2$ & 0.015 p.u., 0.15 p.u., 0.05 p.u.\\
         Line 2 components& $R_3,\,L_3,\,C_3$ & 0.015 p.u., 0.15 p.u., 10 p.u.\\
         \bottomrule
    \end{tabular}
     \label{tab:experiment_parameters}
\end{table}
\renewcommand{\arraystretch}{1}\normalsize

\begin{figure}[b!]
    \centering
    \scalebox{0.478}{\includegraphics[]{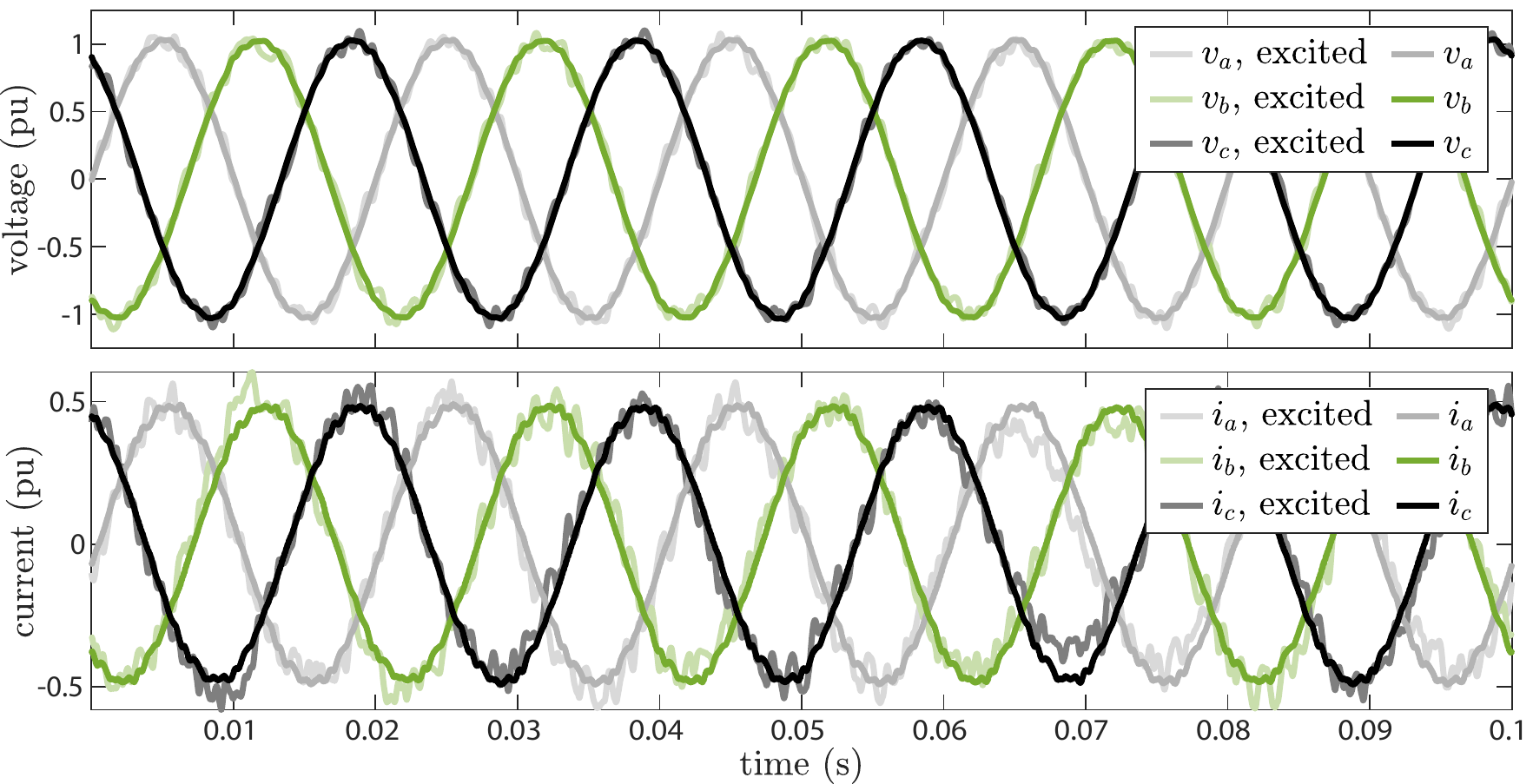}}
    \vspace{-0.4cm}
    \caption{Three phase voltage and current signals at the PCC, both in the absence and in the presence of RBS wideband excitation.}
    \label{fig:signal_comparison}
\end{figure}

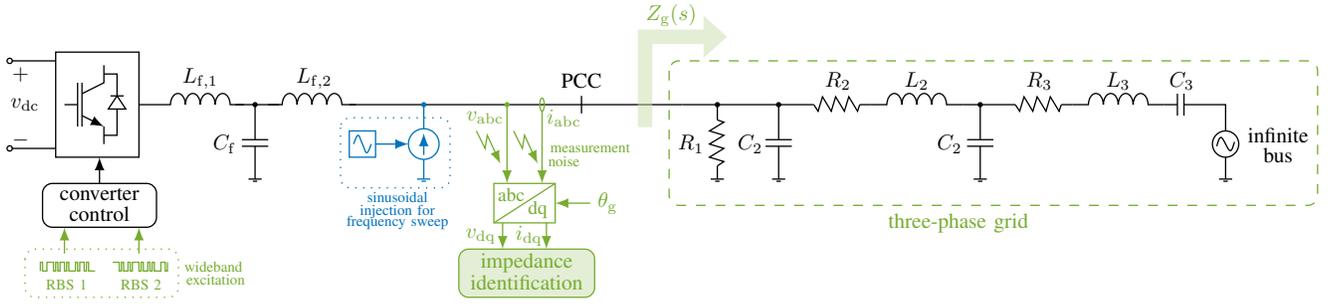
\begin{figure*}[t]
    \centering
    \usetikzlibrary{circuits.ee.IEC}
\usetikzlibrary{arrows}
\tikzstyle{roundnode} =[circle, draw=backgroundcolor2!60, fill=backgroundcolor2!5, scale = 0.5]

\resizebox {1\textwidth} {!} {

\begin{tikzpicture}[circuit ee IEC,scale=0.47, every node/.style={scale=0.65}]

\draw (-2.2,11.8) -- (-1.2,11.8);
\draw (-2.2,9.8) -- (-1.2,9.8);
\draw  (-1.2,12) node (v13) {} rectangle (0.7,9.6);

\draw (-0.1,11.7) -- (-0.1,11.3) -- (-0.6,11.05) node (v3) {};
\draw[-latex](-0.6,10.55) -- (-0.1,10.3) ;
\draw (-0.1,10.3) -- (-0.1,9.9);
\draw (-0.6,11.3) -- (-0.6,10.3);
\draw (-0.7,11.3)--(-0.7,10.3);
\draw (-1,10.8) -- (-0.7,10.8);

\draw (1.1,10.8) node (v7) {} to [inductor={yscale=1.5}] (3.1,10.8) node (v8) {};

\draw (3.65,10.8) node (v7) {} to [inductor={yscale=1.5}] (5.65,10.8) node (v8) {};

\draw (0.7,10.8) -- (1.1,10.8); 
\draw (5.5,10.8) -- (10.75,10.8) -- (13.1,10.8);

\node at (-2,11.5) {$+$};
\node at (-2,10) {$-$};
\node at (-1.9,10.8) {$v_\mathrm{dc}$};

\node at (2.1,11.4) {$L_\mathrm{f,1}$};

\draw(-2.25,11.8) circle (0.6 mm); 

\draw(-2.25,9.8) circle (0.6 mm);

\draw(16.1,10.8) node (v12) {} -- (16.2,10.65) -- (16.3,10.98) -- (16.5,10.65) -- (16.6,10.98) -- (16.8,10.65) -- (16.9,10.98) -- (17.1,10.65) -- (17.15,10.8) node (v15) {};
\draw(20.7,10.8) node (v17) {} -- (20.8,10.65) -- (20.9,10.98) -- (21.1,10.65) -- (21.2,10.98) -- (21.4,10.65) -- (21.5,10.98) -- (21.7,10.65) -- (21.75,10.8);

\draw[rotate=90] (9.4,-13.9) node (v9) {} -- (9.5,-14.05) -- (9.6,-13.72) -- (9.8,-14.05) -- (9.9,-13.72) -- (10.1,-14.05) -- (10.2,-13.72) -- (10.4,-14.05) -- (10.45,-13.9) node (v6) {};

\draw (3.35,10.8) -- (3.35,10);
\fill (3.35,10.8) circle(0.6mm);
\draw (19.6,10) -- (20.2,10); 
\draw (19.6,9.85) -- (20.2,9.85); 
\draw (19.9,9.85) -- (19.9,9.1);
\draw (19.75,9.1) -- (20.05,9.1);
\draw (19.85,9.05) -- (19.95,9.05);
\node at (2.65,9.9) {$C_\mathrm{f}$};

\draw (15,10) -- (15.6,10); 
\draw (15,9.85) -- (15.6,9.85); 
\draw (15.3,9.85) -- (15.3,9.1);
\draw (15.15,9.1) -- (15.45,9.1);
\draw (15.25,9.05) -- (15.35,9.05);

\draw (3.05,10) -- (3.65,10); 
\draw (3.05,9.85) -- (3.65,9.85); 
\draw (3.35,9.85) -- (3.35,9.1);
\draw (3.2,9.1) -- (3.5,9.1);
\draw (3.3,9.05) -- (3.4,9.05);

\fill [color=backgroundcolor](9.1,10.8) circle(0.6mm);

\draw [color=backgroundcolor] (9.9,10.8) ellipse (0.05 and 0.15);
\draw[-latex,color=backgroundcolor] (9.9,10.65) -- (9.9,9);
\draw [-latex,color=backgroundcolor] (9.1,10.8)-- (9.1,9);
\draw [rounded corners = 3, color=backgroundcolor, fill = backgroundcolor!20] (8,7.5) rectangle (11.1,6.4);
\node [color=backgroundcolor] at (9.55,6.75) {identification};
\node  [color=backgroundcolor] at (9.55,7.2) {impedance};
\draw[fill=backgroundcolor!40,color=backgroundcolor!40,opacity=0.5] (12.1,10.3) node (v2) {} -- (12.1,12.5) -- (13.6,12.5) -- (13.6,12.8) -- (14.1,12.35) -- (13.6,11.9) -- (13.6,12.2) -- (12.4,12.2) -- (12.4,10.3) -- (12.1,10.3);
\node [color=backgroundcolor] at (12.85,12.85) {$Z_\mathrm{g}(s)$};

\draw (24.55,10.8) -- (25.5,10.8) -- (25.5,10.2);
\draw  (25.5,9.9) ellipse (0.3 and 0.3);
\draw (25.5,9.6) -- (25.5,9.1);
\draw (25.35,9.1) -- (25.65,9.1);
\draw (25.45,9.05) -- (25.55,9.05);
\draw  [dashed, rounded corners =3,color=backgroundcolor](12.8,11.8) rectangle (27.6,8.5);
\node [color=backgroundcolor] at (19.4,8.1) {three-phase grid};
\draw  plot[smooth, tension=.7] coordinates {(25.3,9.85) (25.4,10.05) (25.6,9.75) (25.7,9.95)};
\draw  [rounded corners=3](1.1,9) rectangle (-1.5,8);
\node at (-0.2,8.75) {converter};
\node at (-0.2,8.3) {control};
\draw [-latex](-0.2,9) -- (-0.2,9.6);
\draw (13.1,10.8) -- (13.9,10.8) node (v10) {}--(13.9,10.45)
;
\draw (13.9,9.4) -- (13.9,9.1);
\draw (13.75,9.1) -- (14.05,9.1);
\draw (13.85,9.05) -- (13.95,9.05);
\draw(13.9,10.8)  -- (15.3,10.8) node (v11) {} -- (15.3,10);
\node at (13.3,9.9) {$R_1$};
\node at (14.65,9.9) {$C_2$};
\draw (17.4,10.8) node (v7) {} to [inductor={yscale=1.5}] (19.5,10.8) node (v8) {};
\draw (15.3,10.8)  --(16.1,10.8); 
\fill (15.3,10.8) circle(0.6mm);
\draw (17.15,10.8) -- (17.4,10.8);
\fill (13.9,10.8) circle(0.6mm);
\draw (19.9,10) -- (19.9,10.8) node (v16) {} -- (19.5,10.8);
\draw (22,10.8) node (v7) {} to [inductor={yscale=1.5}] (24.1,10.8) node (v8) {};
\draw(21.75,10.8) --(22,10.8) ;
\draw (20.7,10.8) --(19.9,10.8);
\draw (24.4,11.05) -- (24.4,10.55);
\draw (24.55,11.05) -- (24.55,10.55);
\fill (19.9,10.8) circle(0.6mm);
\draw (24.1,10.8) -- (24.4,10.8);
\node at (16.65,11.35) {$R_2$};
\node at (18.45,11.35) {$L_2$};
\node at (19.2,9.9) {$C_2$};
\node at (21.25,11.35) {$R_3$};
\node at (23.05,11.35) {$L_3$};
\node at (24.5,11.35) {$C_3$};
\node [backgroundcolor] at (8.6,10.5) {$v_\mathrm{abc}$};
\node [backgroundcolor] at (10.4,10.5) {$i_\mathrm{abc}$};

\draw (10.8,11) -- (10.8,10.6);
\node at (10.8,11.35) {PCC};
\node at (26.7,10.1) {infinite};
\node at (26.7,9.65) {bus};
\draw[scale=0.5,rotate=-180,color=backgroundcolor] (-2.7,-14.4) -- (-2.7,-14) -- (-2.6,-14) -- (-2.6,-14.4) -- (-2.4,-14.4) -- (-2.4,-14) -- (-2.1,-14) -- (-2.1,-14.4) -- (-2,-14.4) -- (-2,-14) -- (-1.9,-14) -- (-1.9,-14.4) -- (-1.8,-14.4) -- (-1.8,-14) -- (-1.6,-14) -- (-1.6,-14.4) -- (-1.5,-14.4) -- (-1.5,-14) -- (-1.4,-14) -- (-1.4,-14.4) -- (-1.1,-14.4) -- (-1.1,-14) -- (-1,-14) -- (-1,-14.4) -- (-0.9,-14.4) -- (-0.9,-14) -- (-0.7,-14) -- (-0.7,-14.4) -- (-0.6,-14.4) -- (-0.6,-14) -- (-0.5,-14) -- (-0.5,-14.4) -- (-0.4,-14.4) -- (-0.2,-14.4);
\draw[scale=0.5,color=backgroundcolor] (-3.1,14) -- (-3.1,14.4) -- (-3,14.4) -- (-3,14) -- (-2.8,14) -- (-2.8,14.4) -- (-2.5,14.4) -- (-2.5,14) -- (-2.4,14) -- (-2.4,14.4) -- (-2.3,14.4) -- (-2.3,14) -- (-2.2,14) -- (-2.2,14.4) -- (-2,14.4) -- (-2,14) -- (-1.9,14) -- (-1.9,14.4) -- (-1.8,14.4) -- (-1.8,14) -- (-1.5,14) -- (-1.5,14.4) -- (-1.4,14.4) -- (-1.4,14) -- (-1.3,14) -- (-1.3,14.4) -- (-1.1,14.4) -- (-1.1,14) -- (-1,14) -- (-1,14.4) -- (-0.9,14.4) -- (-0.9,14) -- (-0.8,14) -- (-0.6,14);
\node at (4.7,11.4) {$L_\mathrm{f,2}$};
\draw (2.9,10.8) -- (3.9,10.8); 
\draw[-latex,color=backgroundcolor] (-1,7.4) -- (-1,8); 
\draw [-latex,color=backgroundcolor](0.7,7.4) -- (0.7,8);
\node [scale = 0.7,color=backgroundcolor] at (-0.95,6.7) {RBS 1};
\node  [scale = 0.7,color=backgroundcolor] at (0.75,6.7) {RBS 2};
\draw (-0.1,10.1) -- (0.2,10.1) -- (0.2,10.7); 
\draw (0.2,11) node (v1) {} -- (0.2,11.5) -- (-0.1,11.5); 
\draw (0.05,11) -- (0.35,11); 
\draw (0.2,11) -- (0,10.7) -- (0.4,10.7) -- (0.2,11);

\draw  [color=backgroundcolor](8.8,9) rectangle (10.2,8.1);
\draw [color=backgroundcolor](8.8,8.1) -- (10.2,9);
\node  [scale=0.9,backgroundcolor] at (9.2,8.75) {abc};
\node  [scale=0.9, backgroundcolor] at (9.8,8.4) {dq};
\draw [-latex,backgroundcolor](9,8.1) -- (9,7.5); 
\draw [-latex,backgroundcolor](10,8.1) -- (10,7.5);

\node [backgroundcolor,scale = 0.7] at (11,9.8) {measurement};
\node [backgroundcolor,scale= 0.7] at (10.4,9.5) {noise};

\draw[-latex,scale=0.5,backgroundcolor] (18.5,20.4) -- (18.9,19.6) -- (19,20.2) -- (19.6,18.9);
\draw[-latex,scale=0.5,backgroundcolor] (16.8,20.4) -- (17.2,19.6) -- (17.3,20.2) -- (17.9,18.9);
\node [backgroundcolor] at (8.5,7.8) {$v_\mathrm{dq}$};
\node [backgroundcolor] at (9.6,7.8) {$i_\mathrm{dq}$};

\draw [-latex,backgroundcolor](11,8.55) -- (10.2,8.55);
\node [backgroundcolor] at (11.4,8.55) {$\theta_\mathrm{g}$};

\draw [rounded corners = 3,dotted,backgroundcolor] (-1.9,7.5) rectangle (1.6,6.4);
\node [backgroundcolor,scale=0.7] at (2.4,7.1) {wideband};
\node [backgroundcolor,scale=0.7] at (2.45,6.8) {excitation};
\fill [backgroundcolor2](7.2,10.8) node (v4) {} circle(0.6mm);
\draw [backgroundcolor2](7.2,10.8)  -- (7.2,10.25);
\draw  [backgroundcolor2](7.2,9.9) ellipse (0.35 and 0.35);
\draw[backgroundcolor2] (7.2,9.55) -- (7.2,9.1); 
\draw[-latex,backgroundcolor2] (7.2,9.65) -- (7.2,10.15);
\draw [backgroundcolor2](7.05,9.1) -- (7.35,9.1); 
\draw[backgroundcolor2] (7.15,9.05) -- (7.25,9.05); 
\draw[backgroundcolor2,-latex] (6.1,9.9) -- (6.85,9.9);
\draw[backgroundcolor2]  (5.5,10.2) rectangle (6.1,9.6);
\draw [backgroundcolor2] plot[smooth, tension=.7] coordinates {(5.6,9.9) (5.7,10.1) (5.8,9.9) (5.9,9.7) (6,9.9)};

\draw [rounded corners = 3, dotted, backgroundcolor2] (5.3,10.5) rectangle (7.8,8.9);
\node[scale=0.7, backgroundcolor2] at (6.6,8.7) {sinusoidal};
\node[scale = 0.7,backgroundcolor2] at (6.6,8.37) {injection for};
\node [scale = 0.7,backgroundcolor2]at (6.6,8.1) {frequency sweep};
\end{tikzpicture}

}
    \vspace{-0.8cm}
    \caption{One-line diagram of the power converter connected to the three-phase power grid used for simulation-based studies of grid impedance identification.}
    \label{fig:simulation_grid}
    \vspace{-0.3cm}
\end{figure*}
\section{Numerical Validation \& Comparison}\label{sec:numerical_experiments}
To validate our proposed parametric grid impedance identification approach and compare it with prevalent nonparametric methods, we use Simscape Electrical in Matlab/Simu- link to perform electromagnetic transients (EMT) simulations based on the test system in \cref{fig:simulation_grid}. We use detailed system and device models, including modulation and the IGBT switching of the VSC. The electrical parameters are provided in \cref{tab:experiment_parameters}. 

\subsection{Validation of the proposed parametric approach} 
During the grid identification experiment, we consider constant stationary grid conditions. To excite the power grid under test, we inject two uncorrelated RBS signals each with an amplitude of 0.1 p.u. and a sampling rate of 5~kHz in the converter's control loop as indicated in \cref{fig:inverter_scheme}. The resulting closed-loop current and voltage responses at the PCC are depicted in \cref{fig:signal_comparison}. We can see that the perturbation level is rather small compared to the non-excited case. Since the excitation is additionally temporally limited to a few seconds, it does not deteriorate the ongoing grid operation.

The voltage and current responses ($v_\mathrm{abc}(k)$ and $i_\mathrm{abc}(k)$) at the PCC are measured (in the presence of measurement noise) at a sampling rate of 5 kHz, and transformed into the $dq$ frame, where the rotating frequency of the $dq$ frame is a constant approximation of the detected 50 Hz grid frequency. The small-signal quantities ($\Delta v_\mathrm{dq}(k)$ and $\Delta i_\mathrm{dq}(k)$) are then computed for parametric system identification. The average variance of the signals' noise (including IGBT switching and measurement noise) is approximately $6.7\cdot 10^{-5}$. 

The true small-signal grid impedance $Z_\mathrm{g}(s)$ of the three-phase system in \cref{fig:simulation_grid} can be analytically computed as a 10th-order continuous $2\hspace{-0.5mm}\times\hspace{-0.5mm}2$ transfer function or state-space system. It serves as a reference model to validate the different system identification techniques in the following. 

Since the recorded voltage and current data set is based on discrete-time samples, the parametric grid impedance model will be (initially) identified in discrete domain. Namely, we compute both an ARX model \cref{eq:arx} and a state-space model \cref{eq:ss} by applying PEM and subspace methods, respectively. This can be carried out by using the Matlab commands ``\texttt{arx}'' and ``\texttt{n4sid}'', which are available in the System Identification Toolbox \cite{ljung2022system}. We also include other processing steps, e.g., prefiltering of the data. If desired, one could additionally consider regularization options (e.g., kernel methods). To get a fair comparison, we chose the same minimal realization order for both model structures. Namely, to compensate for discretization inaccuracies, as well as possible model and disturbance structure mismatches due to the existence of noise in the closed-loop system identification configuration, we iteratively chose to identify slightly higher-order models with a minimal realization order of 16.

\renewcommand{\arraystretch}{1.2}
\begin{table}[b!]\scriptsize
    \centering
    \vspace{-1mm}
           \caption{Comparison of Grid Impedance Identification Methods}
           \vspace{-0.2cm}
               \begin{tabular}{c||c|c|c}
     \toprule
          & \makecell{ Frequency \\ sweep } & \makecell{ Sequential \\ perturbation } & \makecell{ Parametric \\ system identification }  \\ \hline
           \# Experiment cycles & 40 & 2 & 1  \\
         \# Data points & $8\cdot10^6$ & $6\cdot 10^5$ & $3\cdot 10^5$  \\
         Excitation time & 400 s & 30 s & 15 s \\
         Energy of $|\Delta i|$ at PCC & $8.42 \cdot 10^3$ & 829.12 & 412.29  \\
         Energy of $|\Delta v|$ at PCC & $3.29 \cdot 10^3$ & 635.56 & 319.02  \\
         Extra device & yes & no & no \\
         Identified model type &  nonparam. & nonparam. & parametric \\
         Average magnitude error & -31.7 dB & -24.2 dB & -33.6 dB (ARX)\\
        &&& -16.3 dB (n4sid)\\
        Average phase error & 12 deg & 15 deg & 11 deg (ARX)\\
        &&& 33 deg (n4sid)\\
         \bottomrule
    \end{tabular}
     \label{tab:comparison metrics}
\end{table}
\renewcommand{\arraystretch}{1} \normalsize
\begin{figure}[b!]
    \centering
    \scalebox{0.478}{\includegraphics[]{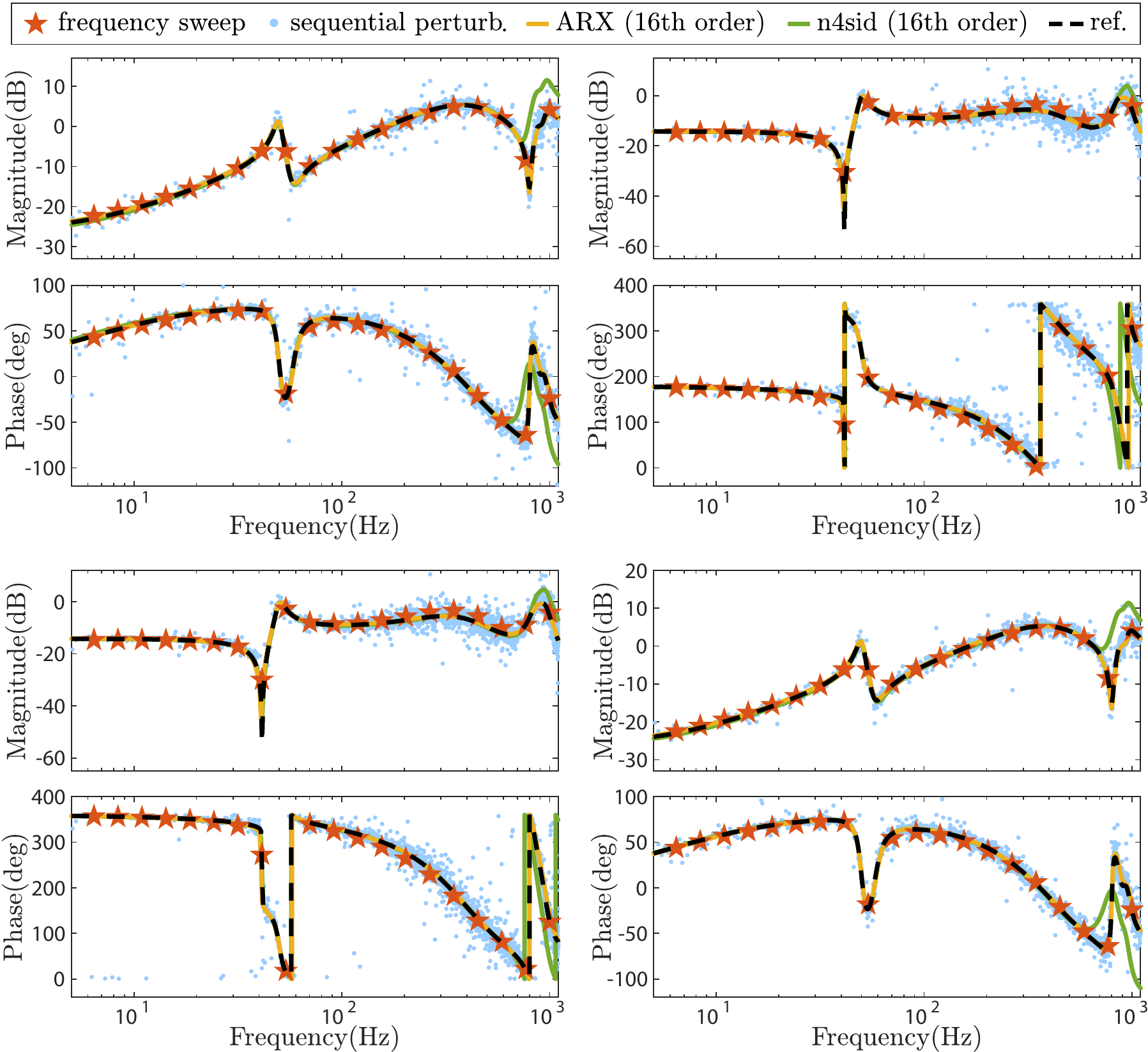}}
    \vspace{-0.4cm}
    \caption{Bode diagrams of the identified $2\hspace{-0.5mm}\times\hspace{-0.5mm}2$ grid impedance $Z_\mathrm{g}(s)$ in \cref{fig:simulation_grid} using different parametric and nonparametric identification methods. The black dashed line indicates the analytically computed true grid impedance. }
    \label{fig:grid_ID_comparison}
\end{figure}
The resulting Bode diagrams of the identified ARX and state-space models are illustrated in \cref{fig:grid_ID_comparison} by the solid yellow and green lines, respectively. We can see that both methods result in a reasonably accurate fitting of the true reference model (indicated by the black dashed line), especially in the frequency range considerably below the Nyquist frequency. In our case, the accuracy of the ARX model is slightly superior at high frequencies, however this may not be general in other system scenarios such as different noise distributions.

\subsection{Comparison to prevalent nonparametric methods}
Next, we demonstrate the competitiveness and superiority of our proposed parametric grid impedance identification approach over the existing nonparametric methods in \cref{fig:mindmap}. To do so, we implement the most commonly used frequency sweep method in \cite{francis2011algorithm,huang2009small}, as well as the sequential perturbation method in \cite{xiao2007novel,shen2013three,xiao2022rapid} (being a representative of wideband excitation methods with low perturbation level) as two representative comparison approaches. Impulse injection methods are not considered here, as they usually do not work very well in practice and are thus rarely applied.

For the sequential perturbation method, we keep the converter configuration as in \cref{fig:inverter_scheme} and perform \textit{two} independent experiment cycles to inject RBS wideband excitation signals in the converter's controller, while measuring the associated current and voltage responses at the PCC. For the frequency sweep method, two independent experiment cycles are conducted for each sinusoidal frequency injection. This results in a total of 40 experiment cycles. The sinusoidal injection mechanism at the PCC is indicated in \cref{fig:simulation_grid} in blue (while keeping the wideband excitation turned off). Having collected the necessary current and voltage data of all experiment cycles appropriately, in both methods, fast Fourier computations are applied to extract harmonic voltage and current responses at each frequency, and pointwise compute the nonparametric grid impedance curve in the frequency domain. Details on the exact computation can be found in the corresponding references on the methods.

Finally, the resulting nonparametric grid impedance curves of the frequency sweep and sequential perturbation method are depicted in \cref{fig:grid_ID_comparison}. It can be seen that, during constant operating conditions of the power grid, they are as accurate as our previously obtained parametric ARX and state-space models, especially in the frequency range considerably below the Nyquist frequency. However, looking at other comparison metrics as listed in \cref{tab:comparison metrics} (e.g., excitation time, number of data points, signal energies, etc.), the superiority of our proposed parametric approach becomes apparent. Moreover, if one was to consider changing operating conditions between the different experiment cycles (e.g., due to rapid load changes or a tripping of a line), the accuracy of the nonparametric methods would considerably deteriorate. The investigation of such scenarios will be part of future work.


\section{Conclusion}\label{sec:conclusion}
We have proposed a parametric approach to MIMO grid impedance identification of three-phase power systems. We inject wideband excitation signals with small perturbation levels in the converter's controller and apply parametric system identification techniques (e.g., PEM or subspace methods) to identify the grid impedance directly from the collected time-domain current and voltage data at the grid interface. Our numerical examples demonstrate the effectiveness of our approach and its superiority to prevalent nonparametric grid impedance identification methods, showing great potential in real-world applications. The fast and accurately identified grid impedance of our approach is well-suited for online assessment of power system stability and adaptive control of grid-connected power converters. Finally, future works should include on-board computations of the identification algorithms, automated order selections, and real-world validations of the proposed approach.

\bibliographystyle{IEEEtran}
\bibliography{IEEEabrv,bibliography}
\end{document}